# Thermodynamic limit on the open circuit voltage of solar cells

Tom Markvart

Engineering Sciences, University of Southampton, Southampton SO17 1BJ, UK, and Centre for Advanced Photovoltaics, Czech Technical University, 166 36 Prague 6, Czech Republic

Abstract: A new thermodynamic limit for the open circuit voltage of solar cells that includes thermalization is obtained in terms of photon entropy. A simple graphical construction makes it possible to link this limit to the existing limits for single junction cells due to Trivich and Flinn, Shockley and Queisser and Würfel, and the thermodynamic Carnot-type limit for hot-carrier solar cell. At the fundamental level, this limit points to similarity between photovoltaic and thermoelectric energy conversion.

With solar cell efficiencies approaching limits to their performance, a careful discussion of these limits becomes increasingly important. Since the maximum current is set by the incident photon flux, the focus is on the open-circuit voltage - in other words, finding the limit on energy conversion efficiency.

The first simple limit of this type,

$$qV_{oc}^{TF} = E_g \qquad (1)$$

where $E_g$ is the semiconductor bandgap, was suggested by Trivich and Flinn in 1955.[1] A more rigorous limit, proposed Shockley and Queisser in 1961,[2] is given by the balance between incident and emitted radiation,

$$qV_{oc}^{SQ} = k_B T_o \ln\left\{\frac{\Phi(T_S)}{\Phi(T_o)}\right\} \qquad (2)$$

with the photon flux density

$$\Phi(T) = 2\pi \int_{\nu_g}^{\infty} \left(\frac{\nu}{c}\right)^2 \frac{d\nu}{e^{h\nu/k_B T} - 1} \qquad (3)$$

where $T = T_S = 6000$ K for solar radiation and $T = T_o = 300$ K for radiation at ambient temperature, $\nu_g = E_g / h$, $h$ is the Planck constant, $k_B$ is the Stefan-Boltzmann constant, and $c$ is the speed of light.

Equation (2) is satisfactory for most purposes but a more general version, valid also for highly concentrated radiation, follows from Würfel's expression for radiation flux with a finite chemical potential:[3]

$$\Phi(T_S) = 2\pi \int_{\nu_g}^{\infty} \left(\frac{\nu}{c}\right)^2 \frac{d\nu}{e^{(h\nu - qV_{oc}^W)/k_B T} - 1} \qquad (4)$$

since, in an ideal solar cell, the electrostatic energy of the carriers separated at contacts is equal to the chemical potential of photons.

These limits – valid for a single-junction cell – should be contrasted with the maximum thermodynamic limit which would be obtained if solar photons with energy $u_S$ in thermal radiation with frequencies above $\nu_g$ were to be converted with Carnot efficiency:

$$qV_{oc}^C = \left(1 - \frac{T_o}{T_S}\right) u_S \qquad (5)$$

Equation (5) applies to a hypothetical device where electron-hole pairs do not lose energy by thermalization – in other words, the hot carrier solar cell. [4,5]

With a view towards next generation photovoltaics, it is of interest to obtain a thermodynamic limit which would include thermalization, and link expressions (1), (2), (4) and (5). Such limit can be obtained by invoking photon entropy and its relationship to the chemical potential.

First, photon flux density $\Phi(T)$ in a beam is transformed into number of photons $N$ in a cavity of volume $V$ that emits this beam. To this end, we can apply an expression for effusion through a hole to radiation gas,[6]

$$\Phi(T) = \frac{c}{4} \frac{N}{V} \qquad (6)$$

We can now define the entropy per photon by

$$s = \left(\frac{\partial S}{\partial N}\right)_{V,T} \qquad (7)$$

where $S$ is the total entropy of radiation in the cavity: blackbody radiation with frequencies extending from $\nu_g$ to $\infty$. Then, using a Maxwell-type relation,[7]

$$\left(\frac{\partial S}{\partial N}\right)_{V,T} = -\left(\frac{\partial \mu}{\partial T}\right)_{V,N} \qquad (8)$$

we integrate (8) with (7) to obtain

$$\mu(T_S) - \mu(T_o) = \int_{T_o}^{T_S} \left(\frac{\partial \mu}{\partial T}\right)_{V,N} dT = -\int_{T_o}^{T_S} s(n_S, T) \, dT \qquad (9)$$

where the photon density $n_S$ in the argument of the entropy $s$ is kept constant at the density corresponding to the solar beam. Noting that the chemical potential $\mu(T_S)$ of solar photons is zero,

$$qV_{oc} = \mu(T_o) = \int_{T_o}^{T_S} s(n_S, T) \, dT \qquad (10)$$

Equation (10), shown in Fig. 1 as an area in the $s$-$T$ plane, is the required general thermodynamic expression that can be used to link the different facets of the open circuit voltage.

The useful energy produced by the solar cell as the integral (10) is equal to the area ① under the curve $s(n_S,T)$. In reality, there are two areas as there are two curves: one corresponding to the rigorous representation of entropy (full line) which gives the open circuit voltage according to Eq. (4). The dashed line is an Sackur-Tetrode-type approximation for the entropy applied to photons:

$$s(n_S, T) = k_B \left\{ \ln\left(\frac{\gamma(T)}{n_S}\right) + T\frac{\gamma'(T)}{\gamma(T)} \right\} \quad (11)$$

where $\gamma(T)$ is the photon density of states (Ref. 6). Equation (11) leads to the Shockley-Queisser value of the open circuit voltage (2).

To proceed further we express the chemical potential in terms of photon energy $u$ and the entropy (7):

$$\mu = u - Ts \quad (12)$$

As the temperature tends to zero, $u \to E_g$, the second term on the right hand side of (12) tends to zero, and the chemical potential tends to the bandgap $E_g$. Hence,

$$\int_0^{T_S} s(n_S, T)\, dT = E_g \quad (13)$$

The integral $\int_0^{T_o} s(n_S, T)\, dT$, represented by the area ②, therefore gives the difference between the Trivich-Flinn voltage (1) and the Shockley-Queisser voltage (2) or, more correctly, the Würfel limit given by (4).

A further analysis is possible by integrating Eq. (10) by parts:

$$qV_{oc} = \left[s(T)T\right]_{T_o}^{T_S} - \int_{T_o}^{T_S} T\, ds = s(T_S)T_S\left(1 - \frac{T_o}{T_S}\right) - \left\{\int_{T_o}^{T_S} T\, ds - \left[s(T_S) - s(T_o)\right]T_o\right\} \quad (14)$$

Clearly, the first term of the right hand side of (14) – represented by the area of the rectangle ①+ ③ in Fig. 1 - corresponds to the Carnot-type voltage (5) as, by (12), the heat $s(T_S)T_S$ is equal to the mean solar photon energy $u_S$ since the chemical potential of the solar radiation is zero. The second term in braces is therefore the energy lost in thermalization, equal to area ③, which has been discussed in Ref. 8, and can be represented by irreversible entropy generation $\sigma_c$.

The integral (10) and Fig. 1 give the open circuit voltage whilst keeping the photon density constant and equal to the solar photon density $n_S$. This is an isochoric process, corresponding to equal étendue of the incident beam $\mathscr{E}_{in}$ and the beam emitted by the solar cell $\mathscr{E}_{out}$, and therefore maximum concentration of sunlight. Further losses arise under one-sun illumination and operation of the cell away from the open circuit. In terms of irreversible entropy generation, these losses are equal to[8]

$$\sigma_{exp} = k_B \ln\left(\frac{\mathcal{E}_{out}}{\mathcal{E}_{in}}\right) \qquad (15)$$

$$\sigma_{kin}(I) = k_B \ln\left(\frac{I_\ell + I_o}{I_\ell + I_o - I}\right) \qquad (16)$$

where $I_\ell$ and $I_o$ are the photogenerated and dark saturation currents. The voltage produced by the solar cell under these more general condition can therefore be written as

$$qV = \int_{T_o}^{T_S} s(T)\,dT - T_o \sigma_{exp} - T_o \sigma_{kin} = \left(1 - \frac{T_o}{T_S}\right) u_S - T_o\left(\sigma_c - \sigma_{exp} - \sigma_{kin}\right) \qquad (17)$$

The fundamental relationship between voltage and entropy extends beyond photovoltaics. Indeed, the voltage generated in a thermoelectric between temperatures $T_\ell$ and $T_h$ is given by

$$V = \int_{T_c}^{T_h} \mathscr{S}\,dT \qquad (18)$$

where $\mathscr{S}$ is the Seebeck coefficient or thermopower (see, for example, Ref. 7). Since $\mathscr{S}$ is the charge-carrier entropy $\mathfrak{s}$ divided by its charge, $\mathscr{S} = \pm \mathfrak{s}/q$, Eq. (18) is just Eq. (10) written for electrons or holes. Vice versa, we could call the photon entropy $s$ divided by the fundamental charge photon thermopower.

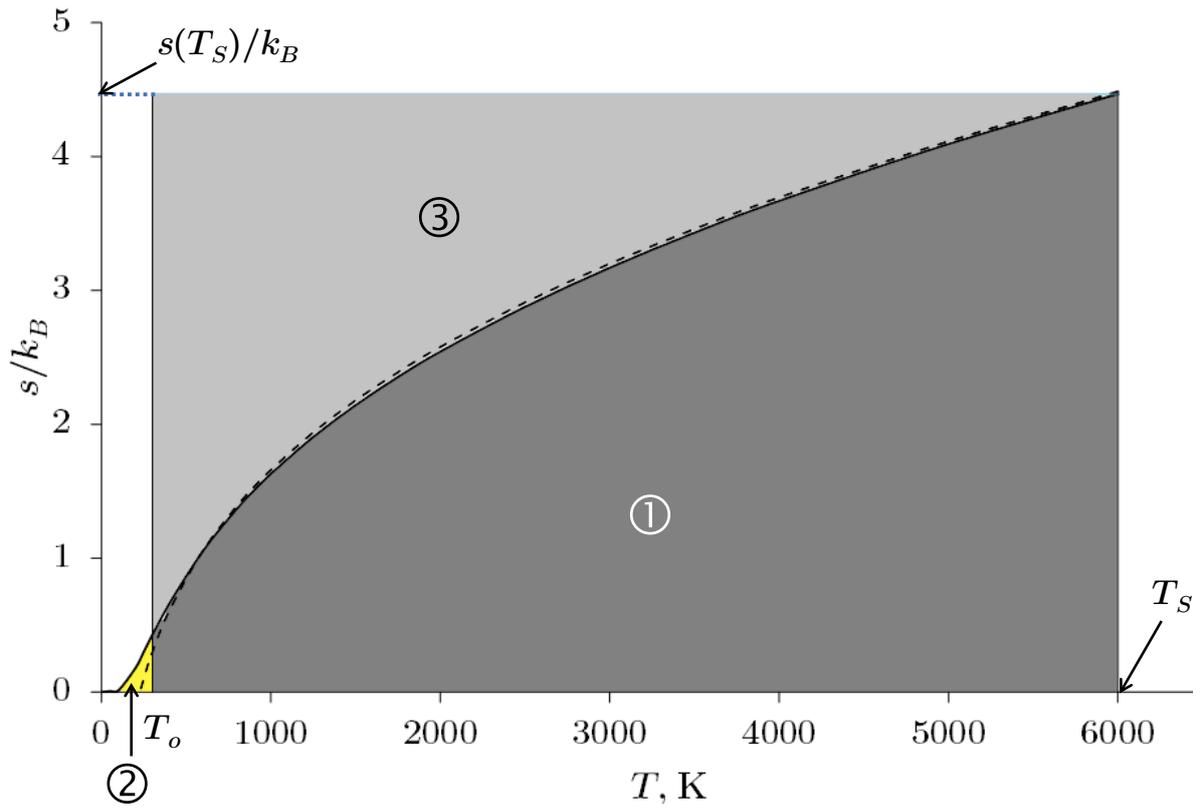

Fig. 1 The entropy *s* per photon as a function of temperature for radiation with photon energies $h\nu > 1.5$ eV. Area ① corresponds to the voltage determined according to the Shockley-Queisser detailed balance (Eq, (2), dashed line) or Würfel's equation (Eq. (4), full line); the sum of areas ① and ② gives the Trivich and Flinn limit (1). Area ③ represents the losses by thermalization, making the rectangle obtained by combining areas ① and ③ equal to the voltage (5) resulting from the conversion of photons with energy $u_S$, with Carnot efficiency.